\documentclass[10pt,preprint,aps,showpacs,nofootinbib]{revtex4}

\usepackage{graphicx}
\usepackage{amsmath}
\usepackage{amssymb}
\usepackage{dcolumn}
\usepackage{bm}
\usepackage{hyperref}

\begin{document}


\title{Nonleptonic charmless two-body $B$ decays involving tensor mesons in the covariant light-front approach}

\author{J. H. Mu\~noz}
\email{jhmunoz@ut.edu.co}
\author{N. Quintero}
\email{nquinte@gmail.com}
\affiliation{Departamento de F\'isica, Universidad del Tolima A. A. 546, Ibagu\'e,
Colombia \vspace{5cm}}

\vspace{7cm}
\begin{abstract}
\vspace{0.3cm}
We reanalyzed  nonleptonic charmless two-body $B$ decays involving tensor mesons in  final
state motivated by the disagreement between current experimental information and theoretical predictions obtained in ISGW2 model for some $\mathcal{B}(B \to P(V)T)$ (where $P$, $V$ and $T$ denote a pseudoscalar, a vector
and a tensor meson, respectively). We have calculated  branching ratios of charmless $B \to PT$ and $B\to VT$ modes, using   $B \to T$ form factors  obtained in the covariant light-front (CLF) approach and the full effective Hamiltonian. We have considered the $\eta-\eta^{\prime }$ two-mixing angle formalism for $B \to \eta^{(\prime)}T$ channels, which increases branching ratios for these processes. Our predictions   obtained in the CLF approach are, in general, greater than those computed in the framework of the ISGW2 model and more favorable with the available experimental data. Specifically, our results for  exclusive channels $B \to \eta K_{2}^{*}(1430)$ and $B \to \phi K_{2}^{*}(1430)$ are in agreement with recent experimental information.
\end{abstract}

\maketitle


\section{Introduction}

In this work we have re-examined  the production of tensor mesons in nonleptonic charmless two-body $B$ decays motivated by the discrepancy between experimental branching ratios for $B \to VT$ and $B \to PT$ modes, reported recently by BaBar Collaboration \cite{BABAR1,BABAR2,BABAR3} and in Particle Data Group (PDG) \cite{PDG}, respectively, and  theoretical predictions obtained in \cite{Kim3} using the ISGW2 model \cite{ISGW2} for evaluating $B \to T$ form factors (see Table I). Keeping in mind that the ISGW2 model presents difficulties in the low-$q^2$ region, we have computed branching ratios for charmless $B \to PT$ and $B \to VT$ modes using the covariant light-front (CLF) approach \cite{Cheng04}, obtaining predictions more favorable with experimental data.\\

There is another interest to study $B \rightarrow VT$ decays is that  they  offer a good scenario in order to investigate about the fraction of longitudinal and transverse decays, similarly to $B \rightarrow V_1V_2$ decays \cite{Chen07,Datta08}. Nowadays, decays with tensor mesons in final state is an area where data is well ahead of the theory. For a recent review about charmless hadronic $B$-meson decays see Ref. \cite{Cheng09}.\\

 At the theoretical level,  there are some works based on quark models that had obtained  branching ratios of nonleptonic two-body $B$ decays including tensor mesons in final state. Initially, Refs. \cite{Katoch,Munoz} calculated at tree level  branching ratios of $B \rightarrow PT$ channels using the nonrelativistic ISGW model \cite{ISGW}. Ref. \cite{Munoz} also computed branching ratios of $B \rightarrow VT$ modes in this model. After, Refs. \cite{Kim1,Kim2}, obtained branchings of charmless $B \rightarrow P(V)T$ channels, using the same model, but considering all the contributions from the   effective Hamiltonian. Ref. \cite{Kim3} is the most recent comprehensive and systematic study  about exclusive charmless $B \rightarrow P(V)T$ decays. In this work, authors  calculated  branching ratios of these modes considering the full effective Hamiltonian and using the improved ISGW2 model \cite{ISGW2} for evaluating  $B \rightarrow T$ form factors. In ISGW2 model, branching ratios are enhanced by about an order of magnitude compared to  previous estimates using ISGW model. Recently, $\mathcal{B}(B^{0} \rightarrow \phi K_{2}^{*0})$ was obtained using the light-front quark model (LFQM) \cite{Chen07}. This prediction is more favorable with experimental data.\\

 At the present,   $B \rightarrow T$ form factors have been calculated in a few quark models: in the ISGW2 model \cite{ISGW2} and in the CLF approach \cite{Cheng04}. Numerical values for  form factors in both models are different. However, the ISGW2 model is not expected to be reliable in the low-$q^2$ region, in particular, at the maximum $q^2=0$ recoil point where the final-state meson could be highly relativistic. In general,  theoretical predictions obtained in Ref. \cite{Kim3} using the ISGW2 model disagree with recent experimental data. In Table I,  we display available experimental branching ratios (see third column) for some exclusive charmless $B \rightarrow P(V)T$ decays and the respective  theoretical predictions reported in Ref. \cite{Kim3} using the ISGW2 model (here $\xi=1/N_c$, where $N_c$ is the color number). We can see that, in general, theoretical predictions are lower that experimental data.\\

 Our aim in this work is to perform a comprehensive and systematic study about $B \rightarrow P(V)T$ decays but taking $B \to T$ form factors from CLF approach \cite{Cheng04}. Additionally, we have included the $\eta-\eta^{\prime }$ two-mixing angle formalism for $B \to \eta^{(\prime)}a_2(K_2^*)$ modes, which increases considerably branching ratios. This mixing has not been considered in previous works \cite{Kim1,Kim3}.

\begin{table}[ht]
{\small Table I.~Comparison between available experimental data for branching ratios (in units of $10^{-6}$) of charmless  $B\rightarrow P(V)T$  decays and  theoretical predictions of Ref. \cite{Kim3} using ISGW2 model}.
\par
\begin{center}
\renewcommand{\arraystretch}{1.3}
\renewcommand{\arrayrulewidth}{0.8pt}
\begin{tabular}{|l|c|c|ccc|}
\hline\hline
  & &  &  & Ref. \cite{Kim3} & \\
  \cline{4-6}
  & Modes &  $\mathcal{B}_{exp}$ & $\xi=0.1$ & $\xi=0.3$ & $\xi=0.5$ \\
  \hline
  & $B^{+} \to \eta K_{2}^{*}(1430)^{+}$  & (9.1 $\pm$ 3.0) \cite{PDG} & 0.256 & 0.031 & 0.028\\
  $B \to PT$ & $B^{+} \to \pi^{+} f_{2}(1270)$  & (8.2 $\pm$ 2.5) \cite{PDG} & 3.284 & 2.874 & 2.491 \\
  & $B^{+} \to K^{+} f_{2}(1270)$ & (9.1$^{+0.4}_{-0.5}$) \cite{PDG}  & 0.394 & 0.344 & 0.298 \\
  & $B^{0} \to \eta K_{2}^{*}(1430)^{0}$   & (9.6 $\pm$ 2.1) \cite{PDG} & 0.237 & 0.029 & 0.026 \\
  \hline
  $B \to VT$ & $B^{0} \to \phi K_{2}^{*}(1430)^{0}$  & (7.8 $\pm$ 1.1 $\pm$ 0.6) \cite{BABAR1} & 0.517 & 2.024 & 4.532 \\
   & $B^{\pm} \to \phi K_{2}^{*}(1430)^{\pm}$  & (8.4 $\pm$ 1.8 $\pm$ 0.9) \cite{BABAR2} & 0.557& 2.18 & 4.881 \\
   & $B^{+} \to \omega K_{2}^{*}(1430)^{+}$  & (21.5 $\pm$ 3.6 $\pm$ 2.4) \cite{BABAR3} & 2.392 & 0.112 & 0.789 \\
  & $B^{0} \to \omega K_{2}^{*}(1430)^{0}$  & (10.1 $\pm$ 2.0  $\pm$ 1.1)\cite{BABAR3}  & 2.221 & 0.104 & 0.732 \\
\hline\hline
\end{tabular}%
\end{center}
\end{table}

This paper is organized as follows: in Sec. II, we discuss about the effective weak Hamiltonian and factorization approach. Sec. III is dedicated to  $B \rightarrow T$ form factors in the CLF approach. In Sec. IV, we present our numerical results and  conclusions are given in Sec. V. In appendix, we show explicitly  expressions for decay amplitudes of $B \to \eta^{(\prime)} a_{2}$ and $B \to \eta^{(\prime)}K_{2}^{*}$ modes, incorporating the $\eta-\eta^{\prime }$ two-mixing angle formalism.\\

\section{Weak Effective Hamiltonian and factorization scheme}

The  effective $\Delta B=1$ weak Hamiltonian $H_{eff}$ for two-body charmless hadronic $B$ decays is \cite{buras96}:

\begin{eqnarray}
H_{eff} &=& \frac{G_F}{\sqrt{2}}\Bigg[V_{ub}V^*_{uq} \Big(C_1(\mu)
O^u_1(\mu) + C_2(\mu) O^u_2(\mu)\Big) + V_{cb}V^*_{cq}\Big(C_1(\mu)
O^c_1(\mu) + C_2(\mu) O^c_2(\mu)\Big)  \nonumber \\
&&- V_{tb}V^*_{tq}\Bigg(\sum^{10}_{i=3}C_i(\mu) O_i(\mu) \Bigg) \Bigg] + h.c.\ ,
\end{eqnarray}

\noindent where $q = d, s$, $G_F$ is the Fermi constant, $C_i(\mu)$ are the Wilson
coefficients evaluated at the renormalization scale $\mu$, and $V_{ij}$  is  the respective Cabibbo-Kobayashi-Maskawa (CKM) matrix element. We show below the local
operators $O_i$ for $b \rightarrow d,s$ transitions:

\begin{itemize}
  \item current-current (tree) operators
  \begin{eqnarray}
  O^u_1 &=& (\bar q_\alpha u_\alpha)_{V-A} \cdot (\bar u_\beta b_\beta)_{V-A}  \nonumber \\
O^u_2 &=& (\bar q_\alpha u_\beta)_{V-A} \cdot (\bar u_\beta b_\alpha)_{V-A}  \nonumber \\
  O^c_1 &=& (\bar q_\alpha c_\alpha)_{V-A} \cdot (\bar c_\beta b_\beta)_{V-A}  \nonumber \\
O^c_2 &=& (\bar q_\alpha c_\beta)_{V-A} \cdot (\bar c_\beta b_\alpha)_{V-A}  \nonumber \\
\end{eqnarray}
  \item QCD penguin operators
  \begin{eqnarray}
  O_{3(5)} &=& (\bar q_\alpha b_\alpha)_{V-A} \cdot \sum_{q^{\prime }}
(\bar q^{\prime }_\beta q^{\prime }_\beta)_{V-A(V+A)}  \nonumber \\
O_{4(6)} &=& (\bar q_\alpha b_\beta)_{V-A} \cdot \sum_{q^{\prime }}
(\bar q^{\prime }_\beta q^{\prime }_\alpha)_{V-A(V+A)}  \nonumber \\
\end{eqnarray}
  \item electroweak penguin operators
  \begin{eqnarray}
  O_{7(9)} &=& \frac{3}{2}(\bar q_\alpha b_\alpha)_{V-A} \cdot
\sum_{q^{\prime }} e_{q^{\prime }}(\bar q^{\prime }_\beta q^{\prime }_\beta)_{V+A(V-A)} \nonumber \\
O_{8(10)} &=& \frac{3}{2}(\bar q_\alpha b_\beta)_{V-A} \cdot
\sum_{q^{\prime }} e_{q^{\prime }} (\bar q^{\prime }_\beta q^{\prime }_\alpha)_{V+A(V-A)},
\end{eqnarray}
\end{itemize}

\noindent where $(\bar q_{1}q_{2})_{V\mp A} \equiv \bar q_{1}\gamma_\mu(1\mp\gamma_5)q_{2}$, and $\alpha$ and $\beta$ are $SU(3)$ color indices. The
sums run over the active quarks at the scale $\mu=\mathcal{O}(m_b)$, i.e. $%
q^{\prime } =u, d, s, c$. \\

In order to obtain  branching ratios of two-body nonleptonic $B \rightarrow M_1M_2$ decays it is necessary to evaluate the hadronic matrix element involving four-quark operators $\langle M_{1}M_{2}|\mathcal{H}_{eff}|B \rangle$. In the framework of factorization approach, it can be expressed as the product of two matrix elements of single currents, which are governed by decay constants and form factors.  The hadronic matrix element is renormalization scheme and scale independent \cite{Buras} while the Wilson Coefficients are renormalization scheme and scale dependent.   \\

For solving the aforementioned scale problem, Refs. \cite{Ali98, Chen99} proposed to extract the $\mu$ dependence from the matrix element $\langle O_i(\mu) \rangle$ and combine it with the $\mu$-dependent Wilson coefficients $C_i(\mu)$ to form $\mu$-independent effective Wilson coefficients $c_i^{eff}$. We have taken  numerical values for them reported in Table I of Ref. \cite{Calderon}. They were calculated using the naive dimensional regularization scheme. \\

It is known that the effective Wilson coefficients $c_i^{eff}$ appear in the factorizable decay amplitudes  as linear combinations. It allows to define the effective coefficients $a_i$, which are renormalization scale and scheme independent, expressed by

\begin{eqnarray}
a_i &\equiv& c^{eff}_i + \frac{1}{N_c} c^{eff}_{i+1}\ (i=odd),  \nonumber \\
a_i &\equiv& c^{eff}_i + \frac{1}{N_c} c^{eff}_{i-1}\ (i=even)\ ,
\end{eqnarray}

\noindent where the index $i$ runs over ($1,...,10$) and $N_c=3$ is the
 number of colors. Phenomenologically,   nonfactorizable contributions to the hadronic matrix element are modeled by treating $N_c$ as a free parameter and its value can be extracted from experiment. In this work we have used  numerical values for  $a_i$ coefficients reported in Table II of Ref. \cite{Calderon}. \\


\section{$B \rightarrow T$ form factors in the CLF approach}

In order to obtain  numerical values of branching ratios of  $B \rightarrow P(V)T$ decays in the framework of generalized factorization, we need to compute the hadronic matrix element $\langle T|J^{\mu}|B \rangle$. We have used the   parametrizations given in  Ref. \cite{ISGW}:

\begin{eqnarray}
  \langle T|V^{\mu}|B\rangle &=& i h(q^{2}) \varepsilon^{\mu\nu\rho\sigma} \epsilon_{\nu\alpha}p_{B}^{\alpha} (p_{B}+p_{T})_{\rho}(p_{B}-p_{T})_{\sigma}, \nonumber\\
  \langle T|A^{\mu}|B\rangle &=& k(q^{2}) \epsilon^{*\mu\nu} (p_{B})_{\nu}+ \epsilon_{\alpha\beta}^{*}p_{B}^{\alpha}p_{B}^{\beta} \nonumber\\
  & & \left[ b_{+}(q^{2})(p_{B}+p_{T})^{\mu} + b_{-}(q^{2})(p_{B}-p_{T})^{\mu} \right],
\end{eqnarray}

\noindent where $V^{\mu}$ and $A^{\mu}$ denote a vector and an axial-vector current, respectively, $\epsilon_{\nu\alpha}$ is the polarization tensor of tensor meson, $p_{B}$ and $p_{T}$ are the momentum of the $B$ meson and the tensor meson, respectively, and  $h, k, b_{\pm}$ are  form factors for the $B \rightarrow T$ transition; $h$ is dimensionless and $k, b_{\pm}$ have dimension of GeV$^{-2}$. \\

At the moment, only two models\footnote{Recently Ref. \cite{LEET} calculated $B \to K_2^*$ form factors  using large energy effective theory (LEET) techniques.}  provide a systematical estimate of $B \rightarrow T$ form factors: the ISGW model \cite{ISGW, ISGW2} and  CLF quark model \cite{Jaus}. Branching ratios for $B \rightarrow P(V)T$ modes using the ISGW2 model were calculated in Ref. \cite{Kim3}. In general, these predictions  present some discrepancies with  experimental data (as illustration, see Table I). Thus, in this work we have used numerical values for form factors $h, k, b_{\pm}$,  obtained in CLF quark model \cite{Cheng04}. This work   has extended the covariant analysis of the light-front approach \cite{Jaus} to even-parity, $p$-wave mesons. \\

 A LFQM can give a relativistic treatment of the movement of the hadron and also provides a fully description of the hadron spin. The light-front wave functions are independent of the hadron momentum and therefore explicitly Lorentz invariant. In the CLF quark model, the spurious contribution, which  depends on the orientation of the light-front, is cancelled by the inclusion of the zero mode contribution, and becomes irrelevant in the decay constants and the form factor, so that the result is guaranteed to be covariant and more self consistent. Recently, this model has been used in several works: Ref. \cite{Lu(Bc)} investigated about semileptonic decays of $B_c$ meson including $s$-wave and  $p$-wave mesons in  final state; Ref. \cite{Lu(Bc)07} studied  nonleptonic $B_c^- \rightarrow X(3872)\pi^-(K^-)$  modes; Ref. \cite{quarkonium} worked with  two-photon  annihilation $P \rightarrow \gamma \gamma$ and magnetic dipole transition $V \rightarrow P\gamma$ processes for the ground-state heavy quarkonium within the CLF approach; Ref. \cite{Radiative} investigated about radiative $B \rightarrow (K^*, K_1, K_2^*)\gamma$ channels in the same framework; and Ref. \cite{Chen07} examined $B \rightarrow (K_{0}^{*}(1430), K_{2}^{*}(1430))\phi$ in the LFQM. In general,  predictions in these works are more favorable with available experimental data.\\

 In CLF approach form factors are explicit functions of $q^2$ in the space-like region and then analytically extend them to the time-like region in order to determine physical form factors at  $q^{2}\geq 0$.   They are parametrized and reproduced in the three-parameter form \cite{Cheng04}:

\begin{equation}\label{1}
F(q^{2}) = \frac{F(0)}{1-aX+bX^{2}},
\end{equation}

\noindent whit $X=q^{2}/m_{B}^{2}$.  Parameters $a$, $b$ and $F(0)$ (form factor at the zero momentum transfer) for $B \rightarrow a_2(1320)$ and  $B \rightarrow K_2^*(1430)$ transitions, which are $B \rightarrow T$ transitions required in this work,  are displayed in Tables VI and VII of Ref. \cite{Cheng04}. In Table II,  we have summarized these numerical values.\\

\begin{table}[ht]
{\small Table II.~Form factors for $B \rightarrow a_{2}(1320)$ and $B \rightarrow K_{2}^{*}(1430)$ transitions obtained in the CLF model \cite{Cheng04}  are fitted to the 3-parameter form in Eq.(\ref{1}).}
\par
\begin{center}
\renewcommand{\arraystretch}{1.2}
\renewcommand{\arrayrulewidth}{0.8pt}
\begin{tabular}{l|ccc|ccc}
\hline\hline
 & & $B \rightarrow a_{2}$ & & & $B \rightarrow K_{2}^{*}$ &\\
 \hline
$F$ & $F(0)$ & $a$ & $b$ &  $F(0)$ & $a$ & $b$\\
\hline
$h$ &  0.008 & 2.20 & 2.30 &   0.008 & 2.17 & 2.22\\
   $k$ &  0.031 & $-2.47$ & $2.47$ &   0.015 & $-3.70$ & 1.78\\
   $b_{+}$ &  $-0.005$ & 1.95 & 1.80 &   $-0.006$ & 1.96 & 1.79\\
   $b_{-}$ &  0.0016 & $-0.23$ & 1.18 &  0.002 & 0.38 & 0.92\\
\hline\hline
\end{tabular}%
\end{center}
\end{table}

Model predictions for $B \rightarrow T$ form factors in the CLF quark model \cite{Cheng04} are different from those in the improved version of ISGW model \cite{ISGW2}. Form factors at small $q^2$ obtained in the CLF and ISGW2 models agree within $40\%$ \cite{Cheng04}. However, when $q^2$ increases $h(q^2)$, $|b_+(q^2)|$ and $b_{-}(q^2)$ increase more rapidly in the light-front model than those in the ISGW2 model \cite{Cheng04}. Another important fact is that the behavior of the form factor $k$ in both models is different (see Table II of Ref. \cite{Chen07}): specifically, for $B \rightarrow K_2^*\phi$, $k(m^2_{\phi})$ is bigger in ISGW2 model than in LFQM: $[k(m^2_{\phi})|_{\text{ISGW2}}]/[k(m^2_{\phi})|_{\text{LFQM}}] = 16.69$.\\

On the other hand, we also need to evaluate  the matrix element of the current between the vacuum and final pseudoscalar $(P)$ or vector $(V)$ mesons. It  can be expressed in terms of the respective decay constants $f_{P(V)}$, in the form

\begin{eqnarray}
\langle P(p_P)|A_\mu| 0\rangle &=& if_{P} q_\mu \nonumber \\
\langle V(p_V,\epsilon)|V_\mu| 0\rangle &=& f_{V}m_{V} \epsilon_\mu,
\end{eqnarray}

\noindent where $q_\mu=(p_{B}-p_{T})_{\mu}$ and $\epsilon_\mu$ is the vector polarization of $V$ meson. Finally, it is important to note that the polarization tensor $\epsilon_{\mu\nu}$ of a $^{3}P_{2}$ tensor meson satisfies the relations

\begin{equation}\label{}
\epsilon_{\mu\nu}=\epsilon_{\nu\mu}, \qquad \epsilon_{\mu}^{\mu}=0, \qquad p_{\mu}\epsilon^{\mu\nu}=p_{\nu}\epsilon^{\mu\nu}=0.
\end{equation}

\noindent Therefore,

\begin{equation}\label{}
\langle 0|(V-A)_{\mu}|T\rangle = a\epsilon_{\mu\nu}p^{\nu}+ b \epsilon^{\nu}_{\nu}p_{\mu}=0,
\end{equation}

\noindent and hence the decay constant of the tensor meson vanishes, i.e., the tensor meson can not be produced from the vacuum. Thus, decay amplitudes for $B \rightarrow PT,\; VT$ processes can be considerably simplified compared to those for two-body charmless $B$ decays such as $B \rightarrow PP,\; PV$, and $VV$ \cite{Ali98,Chen99}, and $B \rightarrow PA,\; AV$, and $AA$ \cite{Calderon}.\\


\section{Numerical Results}

In this section we present  numerical inputs that are necessary to obtain our predictions, and  numerical values for  branching ratios of charmless $B \rightarrow PT$ and $B \rightarrow VT$ decays, using $B \to T$ form factors obtained in the CLF approach \cite{Cheng04}. We used the following values of  decay constants (in GeV units): $f_{\pi}=\text{0.1307}$ and $f_{K}=\text{0.160}$ for pseudoscalar mesons and $f_{\rho}= \text{0.216}$, $f_{\omega}=\text{0.195}$, $f_{\phi}=\text{0.236}$ and $f_{K^{*}}=\text{0.221}$ for vector mesons \cite{PDG}. For  decay constants of $\eta$ and $\eta^{\prime }$ mesons we adopt the $\eta-\eta^{\prime }$ two-mixing angle formalism presented
in \cite{leutwyler, feldmann}, which defines  physical states $\eta$ and $%
\eta^{\prime }$ in function of flavor octet and singlet, $\eta_8$ and $\eta_0$,
respectively:
\begin{eqnarray}
&& |\eta \rangle = \cos \theta |\eta_8\rangle - \sin \theta
|\eta_0\rangle,  \nonumber \\
&& |\eta' \rangle = \sin \theta |\eta_8\rangle + \cos \theta
|\eta_0\rangle.
\end{eqnarray}

\noindent  Decay constants for $\eta_8$ and $\eta_0$ are given by $\langle
0|A^8_\mu|{\eta_{8}}\rangle = i f_8 p_\mu
$ and $\langle 0|A^0_\mu|{\eta_0}\rangle = i f_0 p_\mu$. Assuming that $\eta_8$ and $\eta_0$  are
\begin{eqnarray}
&&|\eta_8 \rangle = \frac{1}{\sqrt{6}} |\bar u u + \bar d d - 2\bar s
s\rangle,  \nonumber \\
&& |\eta_0 \rangle = \frac{1}{\sqrt{3}} |\bar u u + \bar d d + \bar s
s\rangle,
\end{eqnarray}

\noindent they induce a two-mixing angle in the decay constants $f^q_{{%
\eta^{(\prime)}}}$, defined by $\langle 0|\bar q \gamma_\mu \gamma_5 q| {%
\eta^{(\prime)}}(p)\rangle = i f^q_{{\eta^{(\prime)}}}p_\mu$:
\begin{eqnarray}
&& f^u_{\eta^{\prime }} = \frac{f_8}{\sqrt{6}}\sin \theta_8 + \frac{f_0}{%
\sqrt{3}}\cos \theta_0,  \nonumber \\
&& f^s_{\eta^{\prime }} = -2\frac{f_8}{\sqrt{6}}\sin \theta_8 + \frac{f_0}{%
\sqrt{3}}\cos \theta_0  \label{dc1},
\end{eqnarray}

\noindent for the $\eta'$ meson and
\begin{eqnarray}
&& f^u_{\eta} = \frac{f_8}{\sqrt{6}}\cos \theta_8 - \frac{f_0}{\sqrt{3}}\sin
\theta_0,  \nonumber \\
&& f^s_{\eta} = -2\frac{f_8}{\sqrt{6}}\cos \theta_8 - \frac{f_0}{\sqrt{3}}%
\sin \theta_0,
\label{dc2}
\end{eqnarray}

\noindent for the $\eta$ meson. From a complete phenomenological fit of the $\eta-\eta^{\prime }$ mixing
parameters in Ref. \cite{feldmann}, we take $\theta_8=-21.1^\circ$, $%
\theta_0=-9.2^\circ$, $\theta=-15.4^\circ$, $f_8=165$ MeV and $f_0=153$ MeV.
Using these numerical values in Eqs. (\ref{dc1}) and (\ref{dc2}),  decay constants are $%
f^u_{\eta^{\prime }}=61.8$ MeV, $f^s_{\eta^{\prime }}=138$ MeV, $%
f^u_\eta=76.2$ MeV and $f^s_\eta=-110.5$ MeV. For including the $\eta_c$ in the mixing
framework,  we use  decay constants defined
by $\langle 0| \bar c \gamma_\mu \gamma_5 c |{\eta^{(\prime )}} \rangle =
i f^c_{{\eta^{(\prime)}}}p_\mu$. Ref. \cite{feldmann} obtained
$f^c_\eta = -(2.4\pm 0.2)$ MeV and $f^c_{\eta^{\prime }}=-(6.3\pm 0.6)$ MeV.\\

Masses and average lifetimes of neutral and charged $B$ mesons were taken from \cite{PDG}. The running quark masses are given at the scale $\mu \approx m_b$, since the energy released in $B$
decays is of order $m_b$. We use $m_u(m_b)=3.2$ MeV, $m_d(m_b)=6.4$ MeV, $%
m_s(m_b)=127$ MeV, $m_c(m_b)=0.95$ GeV and $m_b(m_b)=4.34$ GeV (see Ref. \cite{fusaoku}). \\

We use  Wolfenstein parameters $\lambda$, $A$, $\bar\rho$ and $\bar\eta$ \cite{wolfenstein} for parametrizing the CKM matrix:

\begin{equation}
V_{\text{CKM}} = \left(%
\begin{array}{ccc}
1-{\frac{1}{2}}\lambda^2 & \lambda & A\lambda^3(\rho-i\eta) \\
-\lambda & 1-{\frac{1}{2}}\lambda^2 & A\lambda^2 \\
A\lambda^3(1-\rho-i\eta) & -A\lambda^2 & 1%
\end{array}%
\right) + \mathcal{O}(\lambda^4),
\end{equation}

\noindent where $\rho = \bar\rho(1-\lambda^2/2)^{-1}$ and $\eta=\bar\eta(1- \lambda^2/2)^{-1}$. We take  central values from the global fit for  Wolfenstein parameters: $\lambda =$ 0.2257, $A$ = 0.814, $\bar\rho$ = 0.135 and $\bar\eta$ = 0.349 \cite{PDG}. \\

For obtaining  branching ratios, we have taken  expressions for  amplitudes of exclusive charmless $B \rightarrow PT$ ($P \neq \eta^{(\prime)}$) and $B \rightarrow VT$ decays given in  appendices of  Refs. \cite{Kim1} and \cite{Kim2}, respectively. These expressions include all the contributions of $H_{eff}$. We do not consider  decay  amplitudes for  $B \to \eta^{(\prime)} a_{2}$ and $B \to \eta^{(\prime)} K_{2}^{*}$ modes  reported  in Ref. \cite{Kim1}. We have worked with  amplitudes displayed in the appendix, which   include the $\eta-\eta^{\prime }$ two-mixing angle formalism. This mixing increases considerably the respective branching ratios.\\

 Our numerical results for branching ratios of exclusive charmless two-body $B \rightarrow PT$ and $B \rightarrow VT$  decays, in the CLF approach, are listed in Tables III - IV and V, respectively.  Our predictions are compared with the work of Kim, Lee and Oh (KLO) \cite{Kim3}, which evaluated  form factors   using the ISGW2 model. We have taken into account theoretical predictions of \cite{Kim3} with $m_s=100$ MeV, $\xi=1/N_c=0.3$ and $\gamma=65^\circ$ (see third column in Tables III and V, and fourth column in Table IV). In Table IV,  we present our numerical predictions for $B \to \eta(\eta')T$ decays: results in second column are obtained in the CLF approach using the amplitudes of Ref. \cite{Kim1} whereas results in third column are obtained  in the same approach but using  amplitudes showed in the appendix (these expressions include the  $\eta-\eta^{\prime }$ mixing).\\

 We have not considered  $B \to P(V)f_{2}$   and $B \to P(V)f_{2}'$ modes because  Ref. \cite{Cheng04} does not make predictions for $B \to f_{2}$ and $B \to f_{2}'$ transitions. They  do not consider  $f_0$, $f_1$ and $f_2$ mesons because their quark contents lying in the mass region of $1.3-1.7$ GeV \cite{Cheng04}. \\

We have analyzed the dependence of branching ratios for $B \to P(V)T$ about form factors. From expressions for decay widths given in Ref. \cite{Munoz} we can observe that $\mathcal{B}(B \to PT)$ and  $\mathcal{B}(B \to VT)$ are quadratic functions of form factors $k$, $b_{\pm}$, and $h$, $k$, $b_+$, respectively. We have explored how is the behavior of these expressions when one changes smoothly numerical values of form factors. We have found that the strongest dependence of  $\mathcal{B}(B \to P(V)T)$ is with respect to the  form factor $b_+$. It is because of kinematical coefficients of $b_+$ are dominant in these expressions. Thus, the precise value of $b_+$ is an important test for both models (CLF and ISGW2). \\

\begin{table}[ht]
{\small Table III.~Branching ratios (in units of $10^{-6}$) for charmless $B \rightarrow PT$ decays, using form factors obtained in CLF model \cite{Cheng04}. Our predictions are compared with the work of KLO \cite{Kim3}}.
\par
\begin{center}
\renewcommand{\arraystretch}{1.3}
\renewcommand{\arrayrulewidth}{0.8pt}
\begin{tabular}{lcc}
\hline\hline
   Process &  This work &  KLO \cite{Kim3} \\
   \hline
   $B^{+} \rightarrow \pi^{+}a_{2}^{0}$ & 4.38 & 2.6  \\
   $B^{+} \rightarrow \pi^{0}a_{2}^{+}$ &  0.015 & 0.001 \\
   $B^{+} \rightarrow K^{+}a_{2}^{0}$ &  0.39 & 0.311 \\
   $B^{+} \rightarrow \pi^{0}K_{2}^{*+}$ & 0.15 & 0.09  \\
   $B^{+} \rightarrow \bar{K}^{0}K_{2}^{*+}$ & 7.84 $\times 10^{-4}$ & 4 $\times 10^{-5}$ \\
   $B^{+} \rightarrow K^{0}a_{2}^{+}$ & 0.015  & 0.011  \\
   $B^{0} \rightarrow \pi^{+}a_{2}^{-}$ & 8.19 & 4.88  \\
   $B^{0} \rightarrow \pi^{0}a_{2}^{0}$ & 0.007 & 0.0003 \\
   $B^{0} \rightarrow K^{+}a_{2}^{-}$ & 0.73 & 0.584 \\
   $B^{0} \rightarrow \pi^{0}K_{2}^{*0}$ & 0.13 & 0.084 \\
   $B^{0} \rightarrow \bar{K}^{0}K_{2}^{*0}$ & 7.15 $\times 10^{-4}$ & 3 $\times 10^{-5}$  \\
   $B^{0} \rightarrow K^{0} a_{2}^{0}$ & 0.014 & 0.005  \\
\hline\hline
\end{tabular}%
\end{center}
\end{table}

\begin{table}[ht]
{\small Table IV.~Branching ratios (in units of $10^{-6}$) for charmless $B \to \eta^{(\prime)} a_{2}$ and $B \to \eta^{(\prime)} K_{2}^{*}$ decays} without and with $\eta-\eta^{\prime }$ mixing, using form factors from CLF model \cite{Cheng04}.
\par
\begin{center}
\renewcommand{\arraystretch}{1.3}
\renewcommand{\arrayrulewidth}{0.8pt}
\begin{tabular}{lcccc}
\hline\hline
   Process &  Without $\eta-\eta^{\prime }$ mixing &  With  $\eta-\eta^{\prime }$ mixing& KLO \cite{Kim3} & Experiment \cite{PDG} \\
  \hline
   $B^{+} \rightarrow \eta a_{2}^{+}$ & 3.78 & 45.8 & 0.294 & - \\
   $B^{+} \rightarrow \eta' a_{2}^{+}$ & 3.72 & 71.3 & 1.31 & - \\
   $B^{+} \rightarrow \eta K_{2}^{*+}$ & 0.65 & 1.19 & 0.031 & 9.1 $\pm$ 3.0 \\
   $B^{+} \rightarrow \eta' K_{2}^{*+}$ & 2.09 & 2.70 & 1.4 & - \\
   $B^{0} \rightarrow \eta a_{2}^{0}$ & 1.77 & 25.2 & 0.138 & - \\
   $B^{0} \rightarrow \eta' a_{2}^{0}$ & 7.20 & 43.3 & 0.615 & - \\
   $B^{0} \rightarrow \eta K_{2}^{*0}$ & 0.59 & 1.09 & 0.029 & 9.6 $\pm$ 2.1 \\
   $B^{0} \rightarrow \eta' K_{2}^{*0}$ & 1.91 & 2.46 & 1.3 & - \\
\hline\hline
\end{tabular}%
\end{center}
\end{table}

\begin{table}[ht]
{\small Table V.~Branching ratios for charmless $B \rightarrow VT$ decays, using form factors obtained in CLF model \cite{Cheng04}. Our predictions are compared with the work of KLO \cite{Kim3}.}
\par
\begin{center}
\renewcommand{\arraystretch}{1.2}
\renewcommand{\arrayrulewidth}{0.8pt}
\begin{tabular}{lccc}
\hline\hline
   Process &  This work ($10^{-6}$) &  KLO \cite{Kim3} ($10^{-6}$) & Experiment \\
  \hline
  $B^{+} \rightarrow \rho^{+}a_{2}^{0}$ & 19.34 & 7.342 & \\
  $B^{+} \rightarrow \rho^{0}a_{2}^{+}$ & 0.071  & 0.007 & $<$ 7.2 $\times 10^{-4}$ \cite{PDG} \\
  $B^{+} \rightarrow \omega a_{2}^{+}$ & 0.14 & 0.01 &\\
  $B^{+} \rightarrow \phi a_{2}^{+}$ & 0.019 & 0.004 &\\
  $B^{+} \rightarrow K^{*+} a_{2}^{0}$ & 2.80 & 1.852 &\\
  $B^{+} \rightarrow \rho^{0} K_{2}^{*+}$ & 0.74 & 0.253 & $<$ 1.5 $\times 10^{-3}$ \cite{PDG} \\
  $B^{+} \rightarrow \omega K_{2}^{*+}$ & 0.06 & 0.112 & $(21.5 \pm 3.6 \pm 2.4)\times10^{-6}$ \cite{BABAR3} \\
  $B^{+} \rightarrow \phi K_{2}^{*+}$ & 9.24 &  2.18 & (8.4 $\pm$ 1.8 $\pm$ 0.9)$\times10^{-6}$ \cite{BABAR2}\\
  $B^{+} \rightarrow \bar{K}^{*0} K_{2}^{*+}$ & 0.59 & 0.014  &\\
  $B^{+} \rightarrow K^{*0}a_{2}^{+}$ & 8.62 & 4.495  &\\
  $B^{0} \rightarrow \rho^{+} a_{2}^{-}$ & 36.18 & 14.686 &\\
  $B^{0} \rightarrow \rho^{0}a_{2}^{0}$ & 0.03 & 0.003 &\\
  $B^{0} \rightarrow \omega a_{2}^{0}$ & 0.07 & 0.005 &\\
  $B^{0} \rightarrow \phi a_{2}^{0}$ & 0.009 & 0.002 &\\
  $B^{0} \rightarrow K^{*+} a_{2}^{-}$ & 7.25 & 3.477 &\\
  $B^{0} \rightarrow \rho^{0} K_{2}^{*0}$ & 0.68 & 0.235 & $<$ 1.1 $\times 10^{-3}$ \cite{PDG}\\
  $B^{0} \rightarrow \omega K_{2}^{*0}$ & 0.053 & 0.104 & $(10.1 \pm 2.0 \pm 1.1)\times10^{-6}$ \cite{BABAR3}\\
  $B^{0} \rightarrow \phi K_{2}^{*0}$ & 8.51 & 2.024 & (7.8 $\pm$ 1.1 $\pm$ 0.6)$\times10^{-6}$ \cite{BABAR1}\\
  $B^{0} \rightarrow \bar{K}^{*0}K_{2}^{*0}$ & 0.55 & 0.026 & \\
  $B^{0} \rightarrow K^{*0}a_{2}^{0}$ & 4.03 &  2.10 &\\
  \hline\hline
\end{tabular}%
\end{center}
\end{table}

\newpage
\section{Conclusions}

In this work we have re-analyzed  charmless two-body hadronic $B \to PT$ and $B\to VT$ decays using  form factors for $B \to T$ transitions from CLF approach, within the framework of generalized factorization. For $B \to \eta^{(')}T$ decays we have considered the $\eta-\eta^{\prime }$ two-mixing angle formalism which increases the respective branching ratios. Our results are compared with ones obtained using ISGW2 model \cite{Kim3} and with available experimental data. Predictions for exclusive $B \to \eta K_{2}^{*}(1430)$ and $B \to \phi K_{2}^{*}(1430)$ in CLF approach are in agreement with experiment data whereas  results obtained in ISGW2 model are lower than them. Our main conclusions are:

\begin{itemize}
\item In general, our branching ratios predictions using the CLF approach in order to obtain   form factors of $B \to T$ transitions are  greater than  previous work of KLO \cite{Kim3} using ISGW2 model.  Predictions in CLF approach seems to be   more favorable with the available experimental data. Some of this modes with branching ratios $\sim (10^{-5}-10^{-6})$ could be measured at  present asymmetric $B$ factories, BABAR and Belle, as well as at future hadronic $B$ experiments such as BTeV and LHC-b.

  \item Our numerical results for  penguin processes $B^{0} \to \phi K_{2}^{*}(1430)^{0}$ and $B^{+} \to \phi K_{2}^{*}(1430)^{+}$ (see Table V) are in agreement with BABAR results \cite{BABAR1,BABAR2} and with the prediction $\mathcal{B}(B^0 \to \phi K_2^{*0})= 7.0 \times 10^{-6}$ reported recently in the theoretical work   \cite{Chen07}. Predictions obtained in Ref. \cite{Kim3} are lower than these experimental data.

  \item The inclusion of   $\eta-\eta'$ mixing effects in  amplitudes  of $B \to \eta^{(\prime)} a_{2}$ and $B \to \eta^{(\prime)} K_{2}^{*}$ modes, increases considerably their branching ratios (see third column in Table IV), so  $B \to \eta^{(')}T$ decays become important.  Although our predictions for branching ratios of $B^{+,0} \to K_2^{*+,0}\eta$ are lower than experimental data they are more favorable  than those of Ref. \cite{Kim3}, which are too lower. Let us mention that even using  amplitudes of Ref. \cite{Kim1} for these modes but working in the CLF framework, the predictions (see second column in Table IV) are bigger than those obtained in ISGW2 model \cite{Kim3}.

  \item In charmless $B \to \pi(K)T$ modes, the bigger discrepancy between  predictions of both models (CLF and ISGW2) appears in the exclusive channels $B^{+,0} \to \pi^{0}a_{2}^{+,0}$ and $B^{+,0} \to \bar{K}^{0}K_{2}^{*+,0}$. The ratio $[\mathcal{B}(B \to \pi^0(\overline{K^0})a_2(K_2^*))_{\text{CLF}}]/[\mathcal{B}(B \to \pi^0(\overline{K^0})a_2(K_2^*))_{\text{ISGW2}}]$ is $\sim$ (15 - 23).

     \item In charmless $B \to VT$ modes (see Table V), the bigger discrepancy between predictions of both models arises from  exclusive penguin channels $B^{+,0} \to \bar{K}^{*0}K_{2}^{+,0}$. In this case, the branching in the CLF approach is $\sim$ (21 - 42) times the one in ISGW2 model.  On the other hand, branchings of  exclusive  $B^{+,0} \to \rho^{0}(\omega)a_{2}^{+,0}$  modes,  in CLF model, are $\sim$ (10 - 14) times than the ones in ISGW2 model. Branchings of $B \to \omega K_2^*$ in CLF approach are the only lower predictions than those obtained in ISGW2 model. However, predictions in both models are lower that experimental data.

\end{itemize}


\section*{ACKNOWLEDGMENTS}
Authors acknowledge financial support from
{\it Comit\'e Central de Investigaciones} of University of Tolima, and G. Calder\'on (Universidad Aut\'onoma de Coahuila - M\'exico) and C. E. Vera (UNAM) for their valuable suggestions. We also thank to G. L\'opez Castro (CINVESTAV) and G. Toledo S\'anchez (UNAM) for reading the manuscript.\\


\begin{appendix}

\begin{center}
\section*{APPENDIX}
\end{center}

In this appendix, we present  expressions for  decay amplitudes of $B \to \eta^{(\prime)} a_{2}$ and $B \to \eta^{(\prime)} K_{2}^{*}$ modes, including the $\eta-\eta^{\prime }$ mixing formalism. Amplitudes displayed below must be multiplied by $i\; G_F\; \epsilon_{\mu\nu}^{*} p_{B}^{\mu}p_{B}^{\nu}/\sqrt{2} $. They are different of expressions displayed in Ref. \cite{Kim1}.

\begin{align}\label{}
\mathcal{A}(B^{0} \to \eta^{(\prime)} a^{0}_{2})=& f^{u}_{\eta^{(\prime)}} F^{B \to a_{2}}(m_{\eta^{(\prime)}}^{2}) \Bigg\{V_{ub}V_{ud}^{*} a_{2} + V_{cb}V_{cd}^{*} a_{2} \Bigg( \frac{f^{c}_{\eta^{(\prime)}}}{f^{u}_{\eta^{(\prime)}}}\Bigg) \nonumber\\
& -V_{tb}V_{td}^{*}\Bigg[a_{4}+2(a_{3}-a_{5}) + \frac{1}{2}(a_{7}-a_{9}-a_{10})- 2\Bigg(a_{6}-\frac{1}{2}a_{8}\Bigg) R \Bigg(1-\frac{f^{u}_{\eta^{(\prime)}}}{f^{s}_{\eta^{(\prime)}}}\Bigg) \nonumber\\
& + (a_{3}-a_{5}+a_{7}-a_{9})\Bigg( \frac{f^{c}_{\eta^{(\prime)}}}{f^{u}_{\eta^{(\prime)}}}\Bigg) +  (a_{3}-a_{5}-\frac{1}{2}(a_{7}-a_{9}))\Bigg( \frac{f^{s}_{\eta^{(\prime)}}}{f^{u}_{\eta^{(\prime)}}}\Bigg)\Bigg]\Bigg\},
\end{align}

\begin{align}\label{}
\mathcal{A}(B^{0} \to \eta^{(\prime)} K_{2}^{*0})=& f^{u}_{\eta^{(\prime)}} F^{B \to K^{*}_{2}}(m_{\eta^{(\prime)}}^{2}) \Bigg\{V_{ub}V_{us}^{*} a_{2} + V_{cb}V_{cs}^{*} a_{2} \Bigg( \frac{f^{c}_{\eta^{(\prime)}}}{f^{u}_{\eta^{(\prime)}}}\Bigg) \nonumber\\
& -V_{tb}V_{ts}^{*}\Bigg[2(a_{3}-a_{5})+ \frac{1}{2}(a_{7}-a_{9}) + (a_{3}-a_{5}+a_{7}-a_{9})\Bigg( \frac{f^{c}_{\eta^{(\prime)}}}{f^{u}_{\eta^{(\prime)}}}\Bigg) \nonumber\\
&  +  \Bigg [a_{3}+a_{4}-a_{5}-\frac{1}{2}(a_{7}-a_{9}+a_{10}) - 2\Bigg(a_{6}-\frac{1}{2}a_{8}\Bigg) X \Bigg(1-\frac{f^{u}_{\eta^{(\prime)}}}{f^{s}_{\eta^{(\prime)}}}\Bigg) \Bigg] \Bigg] \Bigg( \frac{f^{s}_{\eta^{(\prime)}}}{f^{u}_{\eta^{(\prime)}}}\Bigg) \Bigg\},
\end{align}

\noindent with
\begin{equation}\label{}
R= \frac{m_{\eta^{(\prime)}}^{2}}{2m_{s}(m_{b}+m_{d})}, \qquad X= \frac{m_{\eta^{(\prime)}}^{2}}{2m_{s}(m_{b}+m_{s})},
\end{equation}
\noindent and
\begin{equation}\label{}
F^{B \to T}(m_{\eta^{(\prime)}}^{2}) \equiv k(m_{\eta^{(\prime)}}^{2}) + (m_{B}^{2}-m_{T}^{2})b_{+}(m_{\eta^{(\prime)}}^{2}) + m_{\eta^{(\prime)}}^{2} b_{-}(m_{\eta^{(\prime)}}^{2}),
\end{equation}

\noindent where $T$ stands for $a_{2}$ and $K_{2}^{*}$.\\

The factorized decay amplitudes of $B \to \eta^{(\prime)} a_{2}(K_{2}^{*})$ are obtained from amplitudes of $B \to \eta^{(\prime)}\pi(K)$ (see for example appendix A of Ref. \cite{Ali98}) changing $(a_{6}-a_{8}/2)$, $(a_{7}-a_{9})$ and $1/(m_b - m_q)$ by $-(a_{6}-a_{8}/2)$, $-(a_{7}-a_{9})$ and $1/(m_b + m_q)$, respectively, and keeping in mind that a tensor meson $T$ can not be produced from vacuum in generalized factorization. It implies that $\mathcal{A}(B^{+} \to \eta^{(\prime)} a^{+}_{2})=\sqrt{2} \mathcal{A}(B^{0} \to \eta^{(\prime)} a^{0}_{2})$ and $\mathcal{A}(B^{+} \to \eta^{(\prime)} K_{2}^{*+})=\mathcal{A}(B^{0} \to \eta^{(\prime)} K_{2}^{*0})$.
\end{appendix}



\end{document}